\documentclass[conference]{IEEEtran}
\IEEEoverridecommandlockouts
\usepackage{cite}
\usepackage{amsmath,amssymb,amsfonts}
\usepackage{algorithmic}
\usepackage{graphicx}
\usepackage{textcomp}
\usepackage{xcolor}
\usepackage{amsmath,amsfonts}
\usepackage[font=footnotesize]{caption}
\usepackage{algorithmic}
\usepackage{algorithm}
\usepackage{array}
\usepackage[caption=false,font=normalsize,labelfont=sf,textfont=sf]{subfig}
\usepackage{textcomp}
\usepackage{stfloats}
\usepackage{url}
\usepackage{verbatim}
\usepackage{graphicx}
\usepackage{cite}
\usepackage[colorlinks=true, linkcolor=magenta, citecolor=magenta, urlcolor=magenta]{hyperref}
\usepackage{threeparttable}
\usepackage{longtable}
\usepackage{multirow}
\usepackage{float}
\usepackage{array}
\usepackage{booktabs}
\usepackage{tipa}
\usepackage{supertabular}
\usepackage[export]{adjustbox} 
\usepackage{booktabs}
\usepackage{setspace}
\usepackage{mathrsfs}
\usepackage{amsmath}
\usepackage{array}
\usepackage{amssymb}
\usepackage{amsthm}
\usepackage{microtype}
\usepackage{url}
\usepackage{amsfonts,amssymb}
\usepackage{dsfont}
\usepackage{mathtools}
\newcommand{\LaTeXstyle}[1]{{\textsc{#1}}}
\usepackage{listings}

\usepackage{tgheros} 

\def\BibTeX{{\rm B\kern-.05em{\sc i\kern-.025em b}\kern-.08em
    T\kern-.1667em\lower.7ex\hbox{E}\kern-.125emX}}
\begin{document}

\title{Enabling Large Language Models to Perform Power System Simulations with Previously Unseen Tools: \\ A Case of \LaTeXstyle{Daline}
\thanks{This work was supported by the Swiss National Science Foundation under 221126 (Corresponding author: Mengshuo Jia)}
}

\author{\IEEEauthorblockN{1\textsuperscript{st} Mengshuo Jia}
\IEEEauthorblockA{\textit{Power Systems Laboratory} \\
\textit{ETH Zurich}\\
Zurich, Switzerland \\
jia@eeh.ee.ethz.ch}
\and
\IEEEauthorblockN{2\textsuperscript{nd} Zeyu Cui}
\IEEEauthorblockA{\textit{DAMO Academy} \\
\textit{Alibaba}\\
Beijing, China \\
cuizeyu15@gmail.com}
\and
\IEEEauthorblockN{3\textsuperscript{rd} Gabriela Hug}
\IEEEauthorblockA{\textit{Power Systems Laboratory} \\
\textit{ETH Zurich}\\
Zurich, Switzerland \\
hug@eeh.ee.ethz.ch}
}

\maketitle

\begin{abstract}
    The integration of experiment technologies with large language models (LLMs) is transforming scientific research, leveraging AI capabilities beyond specialized problem-solving to become research assistants for human scientists. In power systems, simulations are essential for research. However, LLMs face significant challenges when used to support power system simulations due to limited pre-existing knowledge and the complexity of power grids. To address this issue, this work proposes a modular framework that integrates expertise from both the power system and LLM domains. This framework enhances LLMs' ability to perform power system simulations on previously unseen tools. Validated using 34 simulation tasks in \LaTeXstyle{Daline}, a (optimal) power flow simulation and linearization toolbox not yet exposed to LLMs, the proposed framework improved GPT-4o’s simulation coding accuracy from 0\% to 96.07\%, also outperforming the ChatGPT-4o web interface’s 33.8\% accuracy (with the entire knowledge base uploaded). These results highlight the potential of LLMs as research assistants in power systems.
\end{abstract}


\section{Introduction}\label{sec:Intro}
\IEEEPARstart{C}{ombining} laboratory automation technologies with large language models (LLMs) enables automated execution of scientific experiments \cite{boiko2023autonomous}. Related advances span the fields of mathematics, chemistry, and clinical research, including mathematical algorithm evolution \cite{romera2024mathematical}, geometry theorem proving \cite{trinh2024solving}, chemical experiment design and execution \cite{boiko2023autonomous}, as well as the development and validation of machine learning approaches for clinical studies \cite{tayebi2024large}. These recent achievements signal a new research paradigm, positioning AI as a research assistant for humans with natural language communication abilities, rather than merely a specialized problem solver as in the past. 

Establishing LLMs as research assistants has significant potential for advancing power system studies, which heavily rely on simulations. To develop LLM-based assistants for power systems, it is crucial to equip LLMs with the ability to perform these simulations, a capability not inherent to LLMs. For instance, even GPT-4 often struggles to create small distribution grids using OpenDSS \cite{bonadia2023potential} or writing code for simple power flow problems \cite{dong2024exploring}. This limitation is evident despite the widely available knowledge on optimal power flow problems. However, existing studies mainly focus on conceptualizing \cite{ding2023exploration}, demonstrating \cite{huang2023large, ding2023exploration}, and evaluating \cite{bonadia2023potential, dong2024exploring} LLMs' capabilities in generating power system simulation codes, rather than systematically developing and enhancing their ability to perform these simulations.

To bridge this gap and resolve the above limitation of LLMs, this paper first argues that establishing simulation capabilities in LLMs requires a modular framework that integrates and coordinates multiple techniques. Beyond explicit elements like (i) prompt engineering to enhance LLM performance \cite{huang2023large} and (ii) retrieval-augmented generation (RAG) to incorporate specific power systems knowledge into LLMs \cite{RAG, ding2023exploration, dong2024exploring}, this framework should also consider often overlooked implicit factors: (iii) the refinement of the simulation toolbox (including automated syntax checking and error reporting, and the architecture of the tool's knowledge base), and (iv) the natural language interactive feedback loop between LLMs and the simulation executor.

Building on this concept, this paper proposes a four-module framework to enable LLMs to perform power systems simulations using a simulation toolkit not previously exposed to LLMs\footnote{Precisely, the training data of LLMs does not include relevant information pertaining to the specific toolkit.}. This framework integrates specialization from both the power system and LLM domains. Subsequently, the proposed framework is applied to the \LaTeXstyle{Daline}\footnote{Centered on power system simulations, \LaTeXstyle{Daline} includes functionalities such as (optimal) power flow data generation, data pollution, data cleaning, data normalization, method selection, method customization, model linearization, model evaluation, and result visualization. It supports a large amount of standard power system cases, 57 power flow linearization methods, and over 300 customizable options. See \url{https://www.shuo.science/daline} for more details. } toolbox \cite{Daline} for validation, as  \LaTeXstyle{Daline} was released after the latest updates of any LLMs tested in this paper. Results show that the proposed framework significantly enhances the simulation performance of LLMs. This improvement is a cumulative effect of incorporating multiple techniques, as presented in the following.

\section{Proposed Modular Framework}

\begin{figure*}[t]
\centering
\includegraphics[width=1\linewidth]{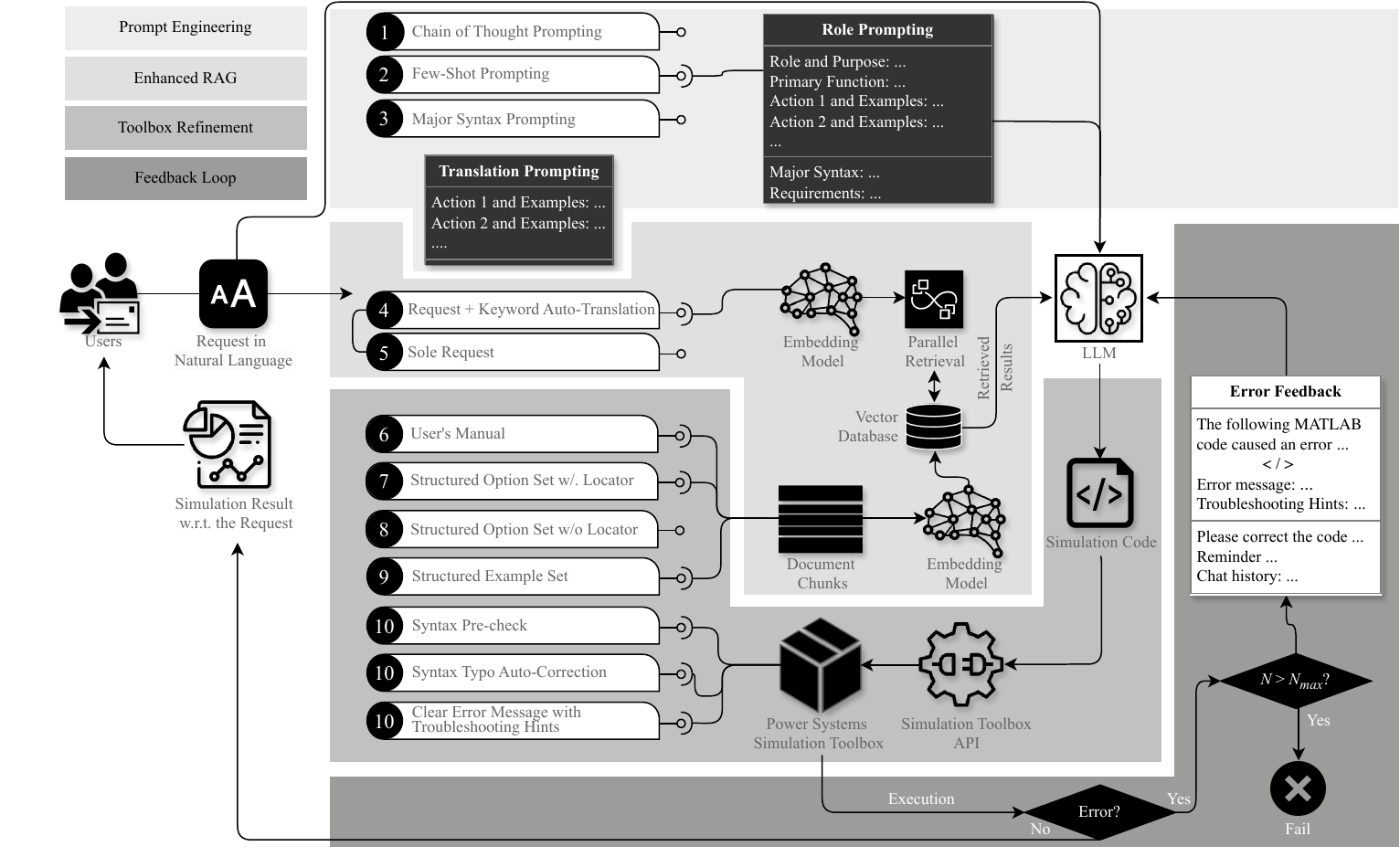} 
\caption{Proposed framework with techniques indexed from 1 to 10. $N$ is the number of feedback iterations and $N_{max}$ is the maximum number of iterations. }
\vspace{-6mm}
\label{fig:framework}
\end{figure*}

The proposed framework consists of four modules with multiple techniques: (i) prompt engineering, (ii) enhanced RAG, (iii) toolbox refinement, and (iv) feedback loop, as illustrated in Fig. \ref{fig:framework} and detailed below. 

\subsection{Prompt Engineering}

To support the LLM to understand its role and purpose, we customized several prompt engineering techniques, including chain of thought prompting \cite{wei2022chain} and few-shot prompting \cite{mann2020language}, for toolbox-based simulations. Beyond clarifying the LLM's role and primary functionality, we defined its actions step-by-step as follows: (i) identifying simulation functions, (ii) syntax learning, (iii) extracting necessary parameters/options, (iv) writing code, (v) providing references, and (vi) drawing conclusions. All steps contain examples for clarity. The major syntax of the toolbox is also explained in the role prompting. While the above prompt engineering techniques mainly originate in the LLM domain, the design of actions, specifics of each prompt, and customization of examples heavily depend on expertise present in the power system simulation tool. For the complete prompt, see the Supporting Document ``\texttt{role\_description.pdf}'' from \href{https://github.com/JarvisETHZ/JarvisETHZ.github.io/blob/master/Supplementary%20File%20for%20Enabling%20Large%20Language%20Models%20to%20Perform%20Power%20System%20Simulations%20with%20Previously%20Unseen%20Tools%20-%20A%20Case%20of%20Daline.zip}{[\underline{here}]}.

\subsection{Enhanced RAG}

For power system simulation tools unfamiliar to LLMs, it is necessary to impart specific knowledge about the tool. RAG \cite{RAG}, a cost-effective approach, can integrate this information into LLMs while reducing hallucinations. Existing studies have used the standard RAG (powered by LangChain) for long-context question answering \cite{dong2024exploring} and non-specific code generation \cite{ding2023exploration} in power systems. The standard RAG procedure includes external knowledge chunking (splitting external documents into smaller pieces), text embedding (converting texts into vectors using existing \texttt{text2vec} neural networks\footnote{The \texttt{text2vec} model we used in this study is from \href{https://help.aliyun.com/zh/dashscope/developer-reference/text-embedding-quick-start?spm=a2c4g.11186623.0.0.5695f97eD8MhdE
}{[\underline{here}]}.}), and information retrieval (finding information in the vector space that matches the whole user request) \cite{dong2024exploring}. However, user requests often involve multiple functions and parameters spread across documents. Simply using the whole request sentence for retrieval may not collect enough semantic information across different sources written at different granularities.

User requests for simulations typically include two critical elements: the functions to be used and the parameters to be set. Hence, to address the above issue, we developed a prompt-based query planning strategy for LLMs. First, we enable LLMs to decompose long requests into sub-requests, each corresponding to a specific simulation function or parameter. Then, we enable LLMs to map each sub-request to a keyword representing the related function or parameter for parallel retrieval. This strategy, leveraging the synergy between LLM and power system simulation expertise, is integrated into the standard RAG structure, resulting in an enhanced RAG architecture that improves the retrieval of critical information from multiple knowledge sources, as shown in Fig. \ref{fig:framework}. The complete query-planning prompt is provided  \href{https://github.com/JarvisETHZ/JarvisETHZ.github.io/blob/master/Supplementary%20File%20for%20Enabling%20Large%20Language%20Models%20to%20Perform%20Power%20System%20Simulations%20with%20Previously%20Unseen%20Tools%20-%20A%20Case%20of%20Daline.zip}{[\underline{here}]}.


\subsection{Toolbox Refinement}

In addition to the previously presented designs, hundreds of tests in our study show that refinement for the simulation toolbox is also needed to reliably enable LLMs to perform power system simulations. This includes (i) developing a RAG-friendly knowledge base, and (ii) a syntax checking and error reporting system, both for the toolbox. 

Specifically, power system simulation toolboxes usually have user manuals detailing functions, parameters, syntax, and examples. While this can be used as external knowledge base for RAG, user manuals are designed for human readability and often spread critical information across different pages, tables, and figures, making them unsuitable for information retrieval. Hence, we propose adding two RAG-friendly documents: one lists all supported parameters/options in the toolbox. Each is written in a separate line, with its name, default value, explanation, and associated functions/methods (acts as a locator to help RAG link parameters with functions/methods). Another contains all code examples from the manual, organized in a predefined structure. These documents help RAG capture more precise information than the user manual alone.

In addition, toolboxes should pre-check syntax and input formats of each function before code execution. Common syntax errors can be corrected internally, while other errors should provide precise messages about the original cause and troubleshooting hints. Although some toolboxes may already have such features, extra attention and further effort are needed when the users are LLMs rather than humans. These features, combined with the feedback loop discussed below, aid LLMs in reasoning and correcting their coding errors automatically.

\subsection{Feedback Loop}

LLMs can make mistakes, but a feedback loop between the simulation executor and LLMs can iteratively correct them. With an established syntax checking and error reporting system, the feedback design amounts to providing a comprehensive error report to LLMs, including (i) the problematic code, (ii) a precise error message, (iii) troubleshooting hints, (iv) a request to correct the code, (v) reminders of common mistakes, and (vi) an organized chat history. This feedback design significantly improves the success rate of LLMs with weaker comprehension abilities, such as GPT-3.5.

\section{Case Study}

{\color{black} In the following, the case study configuration is presented first, followed by an analysis and discussion of the simulation accuracy. }

\begin{table}[t!]
    \scriptsize
    \centering
    \caption{Representative Examples of the Simulation Requests}
    \setlength{\tabcolsep}{5pt} 
    \renewcommand{\arraystretch}{2} 
    \begin{tabular}{m{2cm} p{6cm}} 
    \toprule
    \multicolumn{1}{c}{\textbf{Task Example}} & \multicolumn{1}{c}{\textbf{Simulation Request}} \\
    \midrule
    \textbf{Complex Task 1} & Generate data for 'case9' with 200 training samples and 150 testing samples. Compare and rank the accuracy of the following methods: PLS\_RECW, TAY, the decoupled linearized power flow approach, RR\_KPC, the ordinary least squares method, and the QR decomposition. Set the new data percentage for the method PLS\_RECW to 20\%, and its forgetting factor value as 0.7. Set point0 of the method TAY as 200. For the method RR\_KPC, set the discrete range of tuning eta as logspace(2,5,5), and fix the random seed as 66 for RR\_KPC. Set the response to \{'Vm'\} for all methods. Finally, use the light style for plotting the ranking, and set the type of plotting as 'probability'. Disable the plotting. \\
    \textbf{Normal Task 16} & Generate data for 'case39' with 500 training samples and 250 testing samples. Train a model using LS\_CLS with 5 cross-validation folds and fix the cross-validation partition. \\
    \textbf{Normal Task 20} & Generate data for 'case14' with 400 training samples and 200 testing samples. Compare the accuracy of Decoupled Linearized Power Flow with Data-driven Correction and Power Transfer Distribution Factor for 'case14'. \\
    \textbf{Normal Task 21} & Generate data for 'case39' with 500 training samples and 250 testing samples. Visualize the linearization results for Ridge Regression with the 'academic' theme and disable the plotting. \\
    \bottomrule
    \end{tabular}
\label{table:example}
\end{table}

\begin{table}[t!]
    \scriptsize
    \centering
    \caption{Evaluated Schemes (Technique Index Numbers From Fig. \ref{fig:framework})}
    \setlength{\tabcolsep}{5pt}{ 
    \renewcommand{\arraystretch}{1.6} 
    \begin{tabular}{cccc}
    \toprule
    \textbf{Scheme}         & \textbf{Techniques}       & \textbf{LLM}                      & \textbf{RAG}       \\ \hline
    GPT-4o-Full    & 1,2,3,5,6,7,9,10 & GPT-4o (API)             & Proposed  \\
    GPT-3.5-Full   & 1,2,3,5,6,7,9,10 & GPT-3.5-Turbo (API)      & Proposed  \\
    GPT-3.5-NRPL   & 1,2,3,5,6,8,9,10 & GPT-3.5-Turbo (API)      & Proposed  \\
    GPT-3.5-NRM    & 1,2,3,5,7,9,10   & GPT-3.5-Turbo (API)      & Proposed  \\
    GPT-3.5-NG     & 1,2,3,5,6,7,9    & GPT-3.5-Turbo (API)      & Proposed  \\
    GPT-3.5-NK     & 1,2,3,4,6,7,9,10 & GPT-3.5-Turbo (API)      & Standard \\
    GPT-3.5-NC     & 2,3,5,6,7,9,10   & GPT-3.5-Turbo (API)      & Proposed  \\
    GPT-3.5-NKC    & 2,3,4,6,7,9,10   & GPT-3.5-Turbo (API)      & Standard \\
    GPT-3.5-NRE    & 1,2,3,5,6,7,10   & GPT-3.5-Turbo (API)      & Proposed  \\
    GPT-3.5-NRP    & 1,2,3,5,6,9,10   & GPT-3.5-Turbo (API)      & Proposed  \\
    GPT-3.5-NREP   & 1,2,3,5,6,10     & GPT-3.5-Turbo (API)      & Proposed  \\
    GPT-3.5-NS     & 1,3,5,6,7,9,10   & GPT-3.5-Turbo (API)      & Proposed  \\
    GPT-3.5-Prompt & 1,2,3,10         & GPT-3.5-Turbo (API)      & -      \\
    GPT-3.5-NCS    & 3,5,6,7,9,10     & GPT-3.5-Turbo (API)      & Proposed  \\
    GPT-3.5-NGS    & 1,3,5,6,7,9      & GPT-3.5-Turbo (API)      & Proposed  \\
    GPT-3.5-NKS    & 1,3,4,6,7,9,10   & GPT-3.5-Turbo (API)      & Standard \\
    ChatGPT-4o-R   & 6,7,9,10             & ChatGPT-4o Web Interface & OpenAI   \\
    GPT-4o-R       & 4,6,7,9          & GPT-4o (API)             & Standard \\
    GPT-4o-Sole    & 1,10             & GPT-4o (API)             & -      \\
    GPT-3.5-Sole   & 1,10             & GPT-3.5-Turbo (API)      & -      \\
    \bottomrule
    \end{tabular}}
    \label{table:schemes}
    \end{table}

\subsection{Configuration}

To verify the proposed framework, 34 power system simulation tasks in \LaTeXstyle{Daline} were used for evaluation. These tasks, including 27 normal and 7 complex requests written in natural language, cover the full functionality of \LaTeXstyle{Daline}, from generating AC power flow datasets to data pollution, cleaning, normalization, and power flow linearization. Complex requests also compare and rank the accuracy and computational efficiency of various methods with different settings for training, testing, and visualizing. Each task was tested independently. {\color{black} Table \ref{table:example} provides an overview of the simulation requests by presenting several representative examples of the requests.} The complete set of task requests, as well as all the experiment records and the documents for RAG are available online via this \href{https://github.com/JarvisETHZ/JarvisETHZ.github.io/blob/master/Supplementary%20File%20for%20Enabling%20Large%20Language%20Models%20to%20Perform%20Power%20System%20Simulations%20with%20Previously%20Unseen%20Tools%20-%20A%20Case%20of%20Daline.zip}{[\underline{link}]}.

\begin{figure}[t!]
    \centering
    \includegraphics[width=0.98\linewidth]{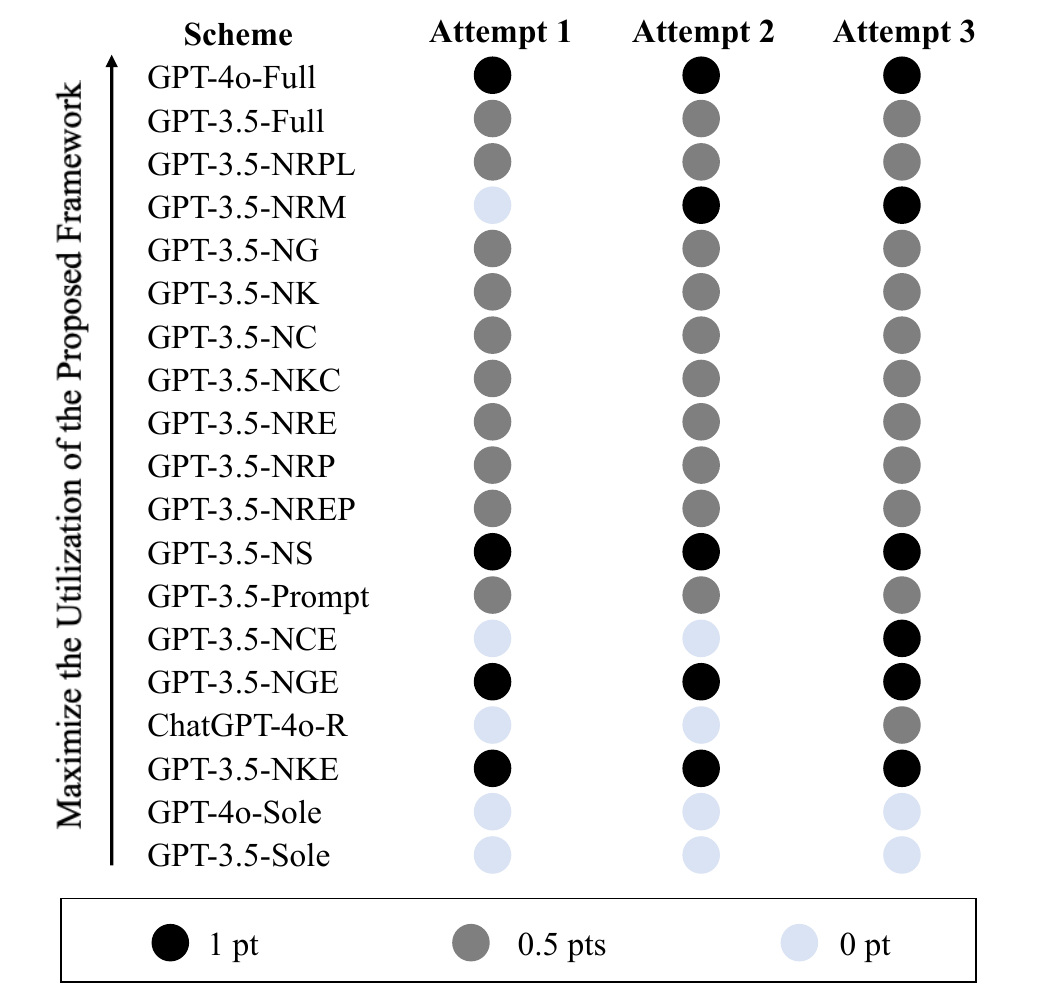} 
    \caption{{\color{black}Score achieved by every scheme in each attempt when processing an example request (i.e., normal task 20, as given in Table \ref{table:example}).}}
    \vspace{-4mm}
    \label{fig:score}
\end{figure}

\begin{figure*}[b!]
    \centering
    \includegraphics[width=0.98\linewidth]{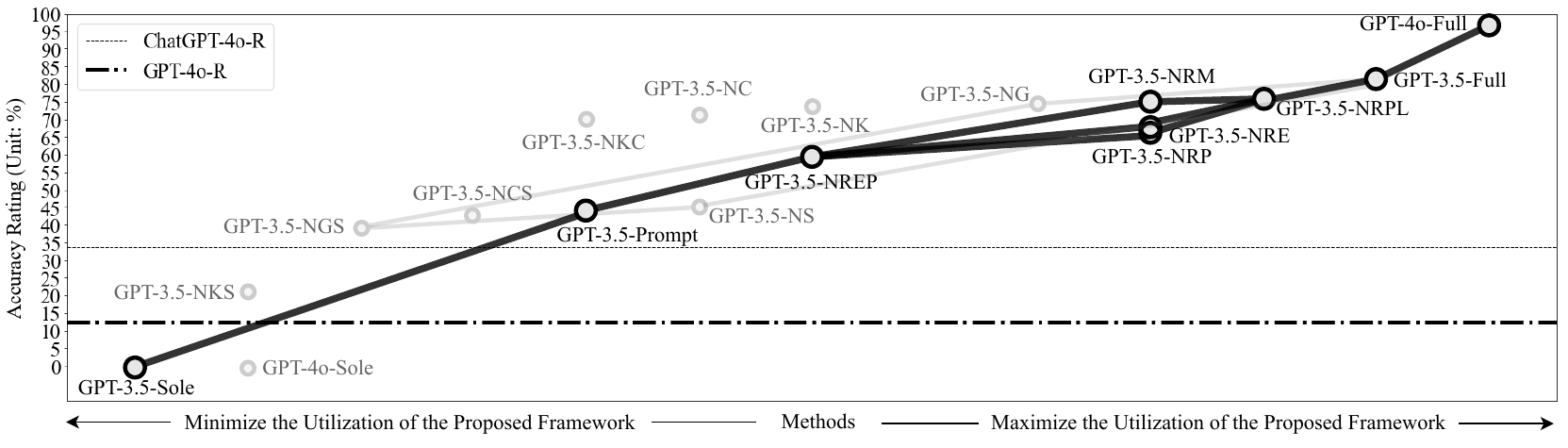} 
    \caption{{\color{black}Overall accuracy of evaluated schemes across both complex and normal tasks (the feedback loop is enabled for all schemes)}.}
    \vspace{-4mm}
    \label{fig:accuracy}
    \end{figure*}    
    For the GPT3.5-NRM scheme, it fails on the first attempt, receiving 0 points for this attempt. However, GPT3.5-NRM automatically corrects its code and successfully addresses the request on the second attempt, earning 1 point for this attempt and an additional point for the following attempt. In contrast, GPT4o-Sole fails all attempts, receiving 0 points for each attempt in response to the simulation request.

    \begin{figure*}[t!]
        \centering
        \includegraphics[width=0.98\linewidth]{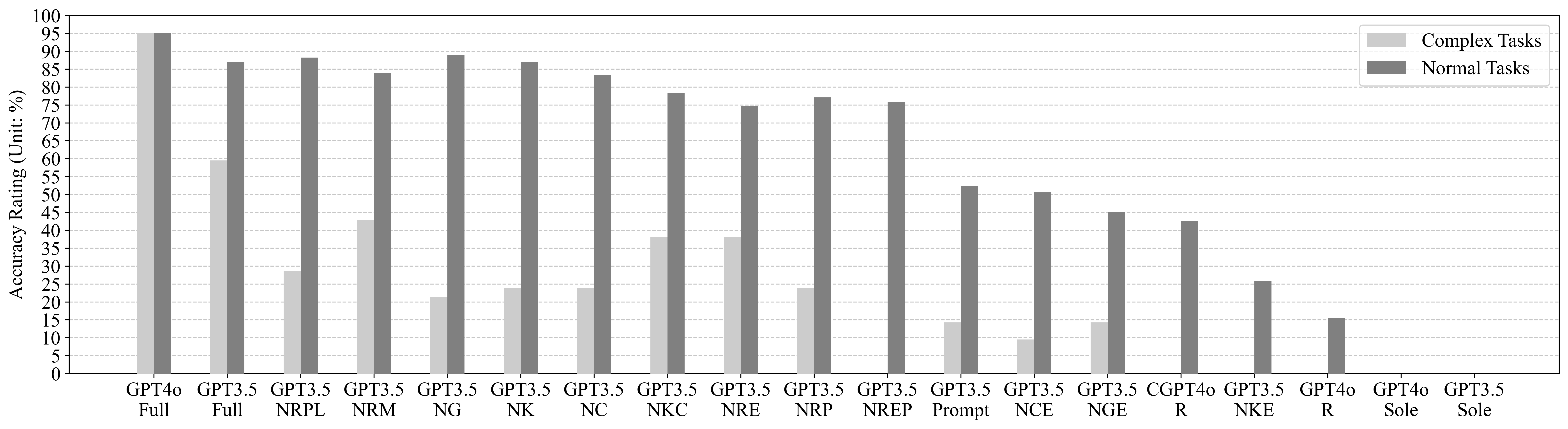} 
        \caption{{\color{black}Individual accuracy of evaluated schemes given the complex or the normal tasks, respectively (the feedback loop is enabled for all schemes)}.}
        \vspace{-4mm}
        \label{fig:bar}
    \end{figure*}

For performance evaluation, 20 schemes listed in Table \ref{table:schemes} were evaluated. Each scheme has 3 attempts ($N_{max}=3$) per simulation request. A scheme earns 1 point per attempt for exact correct code without irrelevant settings, 0.5 points for correct code with irrelevant settings, and 0 points for code with mistakes. Subsequent attempts are made only if the previous one encounters execution errors. Attempts not triggered get the same score as the last attempt. Coding accuracy per scheme is defined as the total points earned divided by the possible highest score ($34 \times 3 = 102$ here), resulting in an accuracy level between 0\% and 100\% per scheme. {\color{black}As an example, Fig. \ref{fig:score} illustrates the score achieved by each evaluated scheme in response to a simulation request (i.e., normal task 20, as given in Table \ref{table:example}). As shown, GPT4o-Full, equipped with the complete version of our proposed framework, successfully completes the simulation request on the first attempt, thereby earning 1 point each for attempts 1, 2, and 3.}

\subsection{Accuracy Analysis and Discussion}

The accuracy performance of the evaluated schemes over all requests is shown in Fig. \ref{fig:accuracy}, {\color{black} and the specific results for each evaluated scheme, categorized by complex and normal tasks, are shown in Fig. \ref{fig:bar}. In the analysis that follows, the accuracy rate refers to the combined accuracy across both complex and normal tasks, unless stated otherwise.}

First of all, both GPT-3.5-Sole and GPT-4o-Sole have zero accuracy, indicating they have not encountered \LaTeXstyle{Daline} before. GPT-4o-R achieves only 12.25\%, suggesting that using the standard RAG only \cite{ding2023exploration,dong2024exploring} is unreliable for LLMs in power system simulations. Even with OpenAI's official RAG tool and the entire knowledge base, ChatGPT-4o-R's accuracy is only 33.82\%. However, with the proposed framework, GPT-4o-Full achieves 96.07\% accuracy. Importantly, the bold black polyline in Fig. \ref{fig:accuracy} shows that incorporating more techniques from the proposed framework significantly improves LLMs' performance.

Additionally, Fig. \ref{fig:accuracy} also highlights the impact of individual techniques on accuracy. For example, the enhanced RAG structure raises accuracy from 74.01\% (GPT-3.5-NK) to 81.37\% (GPT-3.5-Full). Without few-shot prompting, accuracy improves from 20.58\% (GPT-3.5-NKS) to 45.09\% (GPT-3.5-NS) after using the enhanced RAG structure. Once few-shot prompting is implemented, accuracy jumps from 45.09\% (GPT-3.5-NS) to 81.37\% (GPT-3.5-Full). Furthermore, only using RAG-friendly documents as the knowledge base enhances performance (75.49\% accuracy for GPT-3.5-NRM) compared to only using the user manual (60.29\% accuracy for GPT-3.5-NREP). Similarly, syntax error checking and the reporting system combined with few-shot prompting yield significant improvements, as shown by the gray polyline in Fig. \ref{fig:accuracy}. Overall, the accuracy ranking (GPT-3.5-Full $>$ GPT-3.5-NRPL $>$ GPT-3.5-NRM $>$ GPT-3.5-NG $>$ GPT-3.5-NK $>$ GPT-3.5-NC $>$ GPT-3.5-NRE $>$ GPT-3.5-NRP $>$ GPT-3.5-NS) summarizes the contributions of individual techniques. This also demonstrates that achieving high accuracy is a cumulative result of multiple techniques, emphasizing the necessity of a systematic framework with various techniques to enable LLMs to reliably perform complex power system simulations.

{\color{black}
It is also worth noting that complex tasks are generally more challenging for the evaluated schemes, particularly those with a reduced version of the proposed framework, as shown in Fig. \ref{fig:bar}. However, when equipped with the full version of the framework, as in GPT-4o-Full, the scheme achieves a similar, high level of accuracy for both complex and normal tasks. This indicates that the sub-requests within complex tasks are well-identified and managed, comparable to the handling of normal tasks. This result further demonstrates the effectiveness of the proposed framework.}






\section{Conclusion}

This paper proposes a modular framework to enable LLMs to perform power system simulations on previously unseen tools. The framework includes four modules with multiple techniques. Evaluated across 34 different tasks spreading the whole range of capabilities of the \LaTeXstyle{Daline} toolbox, the framework increased coding accuracy for GPT-4o from 0\% to 96.07\%, surpassing the ChatGPT-4o web interface’s 33.82\% accuracy. The impacts of individual techniques have been quantified using 20 different combinations of LLM versions and proposed techniques, demonstrating that high accuracy is achieved through the cumulative effect of multiple techniques. This underscores the necessity of a systematic framework with various techniques to enable LLMs to perform complex power system simulations reliably. Overall, this work highlights the potential for LLMs as research assistants in power systems. Since the proposed framework is currently limited to using a single simulation toolbox, future research will focus on generalizing the framework to accommodate multiple power system simulation tools.

\section*{Acknowledgement}
We would like to acknowledge the assistance of ChatGPT-4o \cite{openai2024chatgpt4o} for language polishing of this paper.

\bibliographystyle{IEEEtran}
\bibliography{IEEEabrv,paper}

\end{document}